\documentclass[aps,nofootinbib,superscriptaddress,twocolumn]{revtex4} % footnotes were getting lost!

\usepackage{amsmath,amssymb,amsfonts}
\usepackage{graphicx}
\usepackage{slashed}
\usepackage{fancyhdr}
\usepackage{subfigure}

\allowdisplaybreaks[1]

\newcommand{\fett}[1]{\mbox{\boldmath$#1$}}

\newcommand{\bqa}{\begin{eqnarray}}
\newcommand{\eqa}{\end{eqnarray}}

\newcommand{\beq}{\begin{equation}}
\newcommand{\eeq}{\end{equation}}
\newcommand{\beqa}{\begin{eqnarray}}
\newcommand{\eeqa}{\end{eqnarray}}

\begin{document}
%%%%%%%%%%%%%%%%%%%%%%%%%%%%%%%%%%%%%%%%%%%%%%%%%%%%%%%%%%%%%%%%%%%%%%%%%%%%%

\renewcommand{\theequation}{\arabic{equation}}

\title{The magnetic form factor of the deuteron in chiral effective field theory}

\author{S.~K\"olling}\email{s.koelling@fz-juelich.de}
\author{E.~Epelbaum}\email{evgeny.epelbaum@rub.de}
\affiliation{Institut f\"ur Theoretische Physik II, Ruhr-Universit\"at Bochum, D-44870, Germany}

\author{D.~R.~Phillips}\email{phillips@phy.ohiou.edu}
\affiliation{Institute of Nuclear and Particle Physics and Department of Phsyics and Astronomy, Ohio University, Athens, OH\, 45701, USA\\}

\begin{abstract}
We calculate the magnetic form factor of the deuteron up to
$O(eP^4)$ in the chiral  EFT expansion of the
electromagnetic current operator. The two LECs 
which enter the two-body part of the isoscalar NN three-current operator
are fit to experimental data, and the resulting values are
of natural size.  The $O(eP^4)$ description of $G_M$ agrees with data for momentum transfers $Q^2 < 0.35$ GeV$^2$.
\end{abstract}


\maketitle

{\it Introduction:} 
In the past two decades chiral effective field theory ($\chi$EFT)
was fruitfully applied to few-nucleon
dynamics (see Refs.~\cite{Epelbaum:2008ga,Epelbaum:2012vx} for recent reviews). Two-nucleon potentials at 
next-to-next-to-next-to-leading order (N$^3$LO) in the chiral expansion 
were developed \cite{Entem:2003ft,Epelbaum:2004fk} which accurately describe  low-energy 
scattering data and the static properties of the deu\-teron. Higher-order corrections to the three-nucleon force are presently under
investigation, see e.g.~\cite{Krebs:2012yv}, 
although discussions regarding non-perturbative renormalization
of the Schr\"odinger equation and implications for the $\chi$EFT power counting continue, see
\cite{Lepage:1997cs,NTvK05,Bi06,PVRA06,EM06} 
for samples of different views. 
 In parallel to these developments in the strong sector, 
much effort has been devoted to pionic and electro\-weak reactions in few-nucleon systems, see
~\cite{Baru:2010xn,Gazit:2008ma,Griesshammer:2012we} 
for recent examples. 

Electromagnetic reactions on light nuclei
such as elastic electron scattering, photo-/electrodi\-sin\-teg\-ration
and radiative capture have been
extensively studied in nuclear physics. In
the single-photon approximation, their theoretical description requires knowledge of the electromagnetic current
operator, which should be constructed consistently with the nuclear
Hamiltonian. The derivation of exchange currents in $\chi$EFT was first addressed in the seminal paper by Park et al., \cite{Park:1995pn}, who,
however, limited themselves to threshold kinematics  $|{\bf q}| \ll M_\pi$ with $M_\pi$ denoting the
pion mass.  Recently, this work was extended to the general
kinematics suitable to study, e.g., electron
scattering off light nuclei at mometum transfer of $|{\bf
  q}|$ of order $M_\pi$ by the JLab-Pisa \cite{Pastore:2008ui,Pastore:2009is,Pastore:2011ip} and Bochum-Bonn groups
\cite{Kolling:2009iq,Kolling:2011mt}. 
Here and in what follows, we discuss the
expansion of the irreducible two-nucleon operators $J_0$ and ${\bf J}$ in powers of $P \equiv (p,m_\pi)/\Lambda$ with $\Lambda$ denoting
the hard scale in the theory, e.g. the cutoff  ($\sim 600$ MeV) used
in calculations. In this expansion the leading-loop order is
$eP^4$. However,  
most of the corrections to the two-body pieces of the two-nucleon current and charge operators at
this order
are of isovector type and thus do not contribute to the
deuteron form factors. In particular, up to this order, the only two-body 
contributions to the isoscalar charge density operator, $J_0^{(s)}$ emerge from the 
leading relativistic corrections of one-pion range so that  $J_0^{(s)}$ is 
parameter free.  
The impact of these corrections on 
the  deuteron charge and quadrupole form factors, $G_C$ and $G_Q$ is studied in
Refs.~\cite{Phillips:2003,Phillips:2006im}.  In these works the deuteron wave functions
obtained from $\chi$EFT potentials at various orders were used to
compute $G_C$ and
$G_Q$ (see also Refs.~\cite{Phillips:1999,Walzl:2001vb} for earlier work along the
same lines). Good agreement with the compilation of elastic
electron-deuteron data from Ref.~\cite{Ab00B} was then found for both form
factors in the kinematic range $Q^2 < 0.35$ GeV$^2$, provided factorization 
was employed in order to account for single-nucleon structure.

On the other hand, the isoscalar two-nucleon current operator, ${\bf J}^{(s)}$ has two two-body
contributions at order $eP^4$: one from a short-distance operator
and one of one-pion range. The impact of these
terms on the magnetic
moments of the deuteron and trinucleons was examined in Ref.~\cite{Song:2007bj}. However, 
markedly more information on the interplay of these terms with each other and with one-body 
mechanisms is available via the 
$|{\bf q}|^2$-dependence of
observables. In this work we present a study in this direction, using $\chi$EFT expressions for ${\bf J}^{(s)}$ derived in Refs.~\cite{Kolling:2009iq,Kolling:2011mt} to 
extend the predictions given for $G_M$ in Refs.~\cite{Phillips:2006im,Walzl:2001vb} to
$O(eP^4)$. We discuss the relevant terms in the current operator and
use the data on the magnetic form factor of the
deuteron at low values of $|{\bf q}|^2$ to determine 
two unknown low-energy constants (LECs).  The $O(eP^4)$ $\chi$EFT 
results thereby obtained accurately describe experimental data on $G_M$ 
in the kinematic range $Q^2 < 0.35$ GeV$^2$. This, together with the findings of Ref.~\cite{Phillips:2006im},
provides a full set of results for elastic electron-deuteron form
factors at $O(eP^4)$. 
 
In the next section we describe
the anatomy of the calculation and outline the relevant terms in the 
two-nucleon current operator. This is followed by a discussion of our
results, including those for the LECs. We finish by summarizing.

{\it Anatomy of the calculation:} The magnetic form factor of the deuteron we are focused in this
work is related to the Breit-frame matrix element of the four-current
operator $J_\mu$ according to the well-known relation
\beq
G_{M} = \frac{1}{\sqrt{2 \eta } | e| }\langle 1 | J^+ | 0 \rangle
\,,
\eeq
where $J^+ = J^1 + i J^2$ and $\eta =| {\bf q} |^2/(4 m_d^2)$ with
${\bf q} \equiv {\bf p}_e ' - {\bf p}_e $ denoting the momentum transfer and $m_d$  the
deuteron mass. (Since we work in the Breit frame we have $Q^2=|{\bf q}|^2$.) The deuteron states are labelled
by the projection of its spin along the direction of $\bf{q}$.  
Both the deuteron wave functions and the current operator appearing on
the right-hand side of the above equation are calculated order-by-order
in $\chi$EFT. 

We now briefly describe the $\chi$EFT expansion of
the two-nucleon current operator, $J_\mu$, as it pertains to the calculation of the deuteron
form factors. We employ Weinberg's power counting throughout this work
which makes use of naive dimensional analysis to determine the
significance  of various contributions. The leading contribution to the charge density, $J_0$, is
given by an $A^0$ photon coupling to a point proton at
order $e$. Nucleon-structure corrections start contributing to the one-body current at
order $eP^2$ [next-to-leading order (NLO)]. The first isoscalar two-body
contribution is generated from  
a tree-level pion-exchange diagram at order
$eP^4$, provided the
nucleon mass is counted as $m_N \sim \Lambda^2/M_\pi \gg
\Lambda$ \cite{Kolling:2009iq,Kolling:2011mt}~\footnote{In the nomenclature of
  Refs.~\cite{Kolling:2009iq,Kolling:2011mt}, these 
  contributions appear at $O(eQ)$.}. This correction is
  associated with a relativistic correction to the one-pion-exchange
  part of the NN potential. There are numerous other
corrections to the two-body part of the charge operator at leading-loop order, $eP^4$, from one- and two-pion
exchange diagrams and from pion loops involving the lowest-order
contact interactions, but this is the only isoscalar effect. The explicit form of all terms can be found in
Refs.~\cite{Kolling:2009iq,Kolling:2011mt}. 

The chiral expansion of the three-current starts at order $eP$ with
the single-nucleon contributions. The first two-body terms emerge
from tree-level one-pion exchange diagrams at order $eP^2$ [NLO].  
The next two-body corrections to ${\bf J}$ occur at order $eP^4$
from pion loops and tree diagrams involving higher-order vertices from 
the effective Lagrangian.  The two-pion exchange contributions are
parameter-free \cite{Kolling:2009iq}, while the one-pion exchange terms depend on
the LECs $\bar d_8$,  $\bar d_9$, $\bar d_{18}$, $\bar d_{21}$ and
$\bar d_{22}$ entering $\mathcal{L}_{\pi N}^{(3)}$\cite{Park:1995pn,Ecker:1995rk,Fettes:1998ud,Gasser:2002am}. However, the only long-range two-body mechanism in ${\bf J}^{(s)}$ at this order is proportional to $\bar{d}_9$. While this LEC could, in principle, be constrained
by pion photoproduction data, in practice these data provide little information on $\bar{d}_9$~\cite{FHLU,Gasparyan:2010xz}.  

$G_M$ is of particular interest at $O(eP^4)$ because it is there that the first short-distance NN physics not determined by NN scattering and gauge invariance appears. This is represented by the simplest M1 isoscalar four-nucleon-one-photon contact term in the $\chi$EFT Lagrangian, which is of the form~\cite{Ch99,Walzl:2001vb,Kolling:2011mt}:
\begin{equation}
{\cal L}_{M1}=\frac{e L_2}{2} \left( N^\dagger \,\epsilon_{ijk} \sigma_i F_{jk} \,N\right)\left(N^\dagger  \, N \right).
\label{eq:LM1NN}
\end{equation}
The low-energy constant (LEC) $L_2$ that appears in Eq.~(\ref{eq:LM1NN}) must be extracted from data on electromagnetic reactions in the two-nucleon system. 

The combination of these two effects yields a two-body isoscalar current operator ${\bf J}^{(s)}$~\cite{Kolling:2009iq,Kolling:2011mt}:
\begin{eqnarray}
{\bf J}_{2B}^{(s)} & = & 2e\frac{g_A\, i}{F_\pi^2}
    d_9 \, \fett{\tau}_1 \cdot\fett{\tau}_2 \,
 \frac{\fett{\sigma}_2\cdot\bf{q}_2}{q_2^2+M_\pi^2}\left[\bf{q}_1
   \times\bf{q}\right] \nonumber \\
&&{}+  i e L_2 \,\left(\fett{\sigma}_1 + \fett{\sigma}_2
  \right) \times \bf{q}_1  + (1 \leftrightarrow 2)\,,
\label{eq:J2Bs}
\end{eqnarray}
where $\bf{q}$ labels the photon momentum and $\bf{q}_{1/2}$ labels the
momentum transfer on nucleon one/two respectively.
Since $G_M$ is determined completely by the one-body part of ${\bf J}^{(s)}$ up to $O(eP^3)$ the total form factor is thus
\begin{equation}
G_M= \frac{1}{\sqrt{2 \eta } | e| }\langle 1|{{\bf J}^{(s)}_{1B}}^+ + {{\bf J}_{2B}^{(s)}}^+|0 \rangle.
\label{eq:GM}
\end{equation}
Here we use factorization to compute ${\bf J}_{1B}$, i.e. we write:
\begin{equation}
{{\bf J}_{1B}^{(s)}}^+=\frac{|e|}{M}[G_E^{(s)}(Q^2) 2 {\bf p}^+ + i G_M^{(s)}(Q^2)  ({\mathbf \sigma}_1 \times {\bf q})^+],
\end{equation}
with ${\bf p}$ the momentum of the struck nucleon, and $G_E^{(s)}$  and $G_M^{(s)}$ the isoscalar single-nucleon form factors, for which we take the parameterization of Ref.~\cite{BHM}. The use of this ansatz for the one-body part of ${{\bf J}^{(s)}}^+$ is equivalent (up to corrections that begin only two orders beyond the order to which we work) to making a $\chi$EFT expansion for the ``body" form factors $D_M$ and $D_E$~\cite{GG02}. This allows us to focus on the momentum transfer at which the $\chi$EFT expansion for the NN current operator ${\bf J}$ breaks down, without having to worry whether the theory is doing a good job of describing isoscalar nucleon structure.

{\it Results:} We now evaluate the matrix elements in Eq.~(\ref{eq:GM}) with a variety of $\chi$EFT deuteron wave functions computed with the NLO and NNLO $\chi$EFT
potentials and different values of the cutoffs $\Lambda$  in the
Lippmann-Schwinger equation and $\tilde{\Lambda}$ in the spectral function. The result found for $G_M$ with LO $\chi$EFT wave functions and the leading piece of ${\bf J}^{(s)}$, denoted here as $O(eP)$, was computed in Ref.~\cite{Pa08}. Corrections to this come both from higher-order pieces of the NN potential, $V$, which affect the wave function, and from the corrections to ${\bf J}^{(s)}$ discussed in the previous section. The NNLO $\chi$EFT 
potential includes all effects up to $O(P^3)$
relative to leading (in this counting), so its deuteron wave function, when combined 
with the $O(eP^4)$ ${\bf J}^{(s)}$, yields a $\chi$EFT calculation
for $G_M$ which includes all effects up to $O(eP^4)$.

The pertinent matrix elements are computed via
Monte-Carlo (MC) integration. 
To increase efficiency, we use importance sampling with the weight function of Ref.~\cite{LIE}:
\begin{equation}
  \label{eq:weightfunction}
    p\left(\bf{k}\, \right) \equiv  p(k) = \frac{(r-3)(r-2)(r-1)}{8\pi}
  \frac{C^{r-3}}{(k + C)^r}\, .
\end{equation}
The functional form of $p(k)$ is chosen such that the weight function
is maximal at the origin, reflecting the large $S$-wave component of
the deuteron wave function. The parameters $C$ and $r$  control the
vanishing of the  weight function at large momenta and are tuned to
optimal values (in terms of the efficiency of the MC
integration) by calculating the expectation value of the one-pion
exchange potential yielding $C=1$ GeV and $r=11$.    

As in Ref.~\cite{LIE} we perform a path average over several runs. We use
2730 sample points and and the path average is performed for 3000 runs. Analysis
of the run-to-run fluctuations indicates a final answer with
better than 1\% precision throughout the momentum range of  $0-800$ MeV.
At several points we compared this MC answer to calculations using quadrature methods,
and always found agreement within the precision claimed. 

We adopt the following procedure to determine the values of the two
LECs entering ${\bf J}^{(s)}$. First, we fix the value of $L_2$ for a
given $\bar{d}_9$ by demanding that the magnetic moment of 
the deuteron is reproduced. We then perform a $\chi^2$-fit to the experimental
data for $|{\bf q} |< 400$ MeV (including four points from the parametrization of Ref.~\cite{SIC}) to
determine $\bar{d}_9$ . Our attempts to use even lower-$|{\bf q}|$
data for this fit resulted in unstable answers,
reflecting the insensitivity of $G_M$ to this LEC at small values of $|{\bf q}|$.

\begin{figure}[tb]
\vskip 0.6 true cm
\begin{center} 
 \includegraphics[width=0.49\textwidth,keepaspectratio,angle=0,clip]{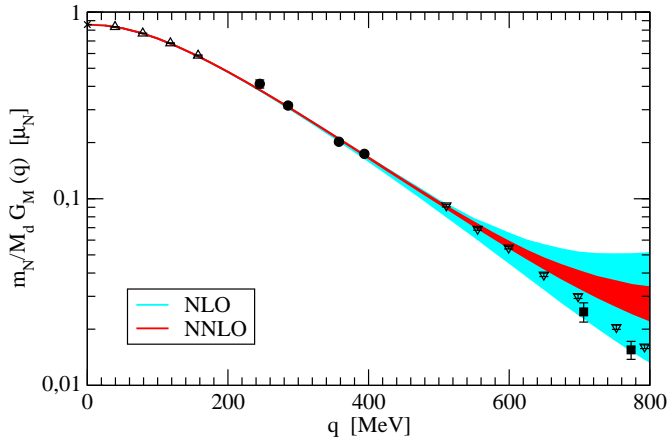}
\end{center} 
\vskip -0.2 true cm
\caption{The magnetic formfactor $G_M$ as a function of $|{\bf q}|$. Experimental
    data for the magnetic moment is from~\cite{LIN}.  The
    remaining data  are from the
    parameterization of~\cite{SIC} (upward triangles) and scattering
    experiments reported in \cite{AUF} (downward triangles),
    \cite{CRA} (squares) and \cite{SIM} (solid dots). The light blue (dark red) band represents the
     results with NLO (NNLO) 
    wave functions, and ${\bf J}^{(s)}$ computed up to $O(eP^4)$. }
  \label{fig:PlotGMfull}
\end{figure}

The results of this procedure are shown in Fig.~\ref{fig:PlotGMfull}. 
The light/blue (dark/red) band
is obtained using wave functions computed with the NLO (NNLO)
$\chi$EFT potential. The width of the band shows the variation of
the prediction as $\Lambda$ and $\tilde{\Lambda}$ are changed in the 
range $\Lambda = 400\ldots 550$ MeV ($\Lambda = 450\ldots 600$ MeV) at NLO (NNLO) and 
$\tilde \Lambda = 500\ldots 700$ MeV.    
The cutoff variation is reduced at NNLO, and the data well described
for $Q^2 < 0.35$ GeV$^2$. 

\begin{figure}[tb]
\vskip 0.6 true cm
\begin{center}
  \includegraphics[width=0.49 \textwidth,keepaspectratio,angle=0,clip]{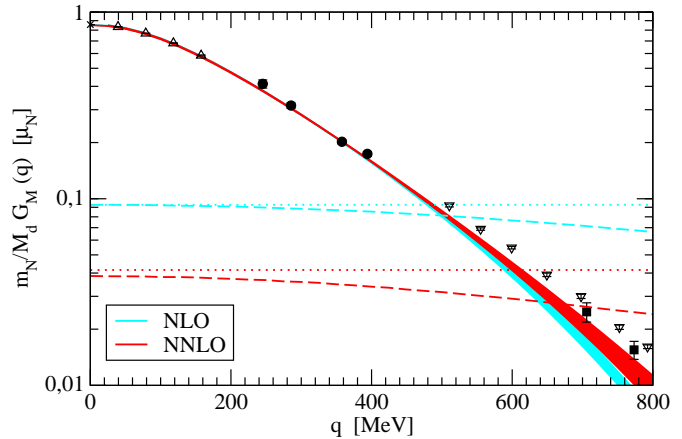}
\end{center}
      \vskip -0.2 true cm
  \caption{The magnetic formfactor $G_M$ as a function of $|{\bf
      q}|$. 
The light blue (dark red) band represents the results with NLO (NNLO)
    wave functions using the impulse approximation. The dashed/dotted lines are the contributions from
    two-body pieces of ${\bf J}^{(s)}$, as described in the text. For
    remaining notation see Fig.~\ref{fig:PlotGMfull}. }
  \label{fig:GMbreakdown} 
\end{figure}

In order to assess the momentum transfer at which the $\chi$EFT expansion for ${\bf J}^{(s)}$ breaks down, in 
Fig.~\ref{fig:GMbreakdown} we 
show the size of different contributions to the final result. This time the bands
represent the impulse approximation result obtained with NLO (light/blue) and NNLO (dark/red) wave functions. 
The dotted (dashed) line is the effect from the piece of ${\bf J}_{2B}^{(s)}$ that is proportional to $L_2$ ($\bar{d}_9$). 
For both two-body matrix elements, we show results averaged over the five cutoff combinations considered, with 
the light blue lines showing the NLO case, and the dark red lines obtained 
with NNLO wave functions. We estimate the breakdown scale of the EFT expansion by values of
momentum transfer at which the $O(eP^4)$ two-body
contributions start becoming comparable to the effect of the $O(eP)$ (impulse-approximation) piece of the current.
Fig.~\ref{fig:GMbreakdown} shows that the smaller two-body contributions to $G_M$ found with the NNLO wave function delay the breakdown of the
expansion. Even so, we would infer a breakdown scale $|{\bf q}|=600$ MeV, as there the short-distance effect $\sim L_2$ becomes equal in magnitude
to the impulse-approximation result. 

\begin{table}[hb]
  \centering
   \begin{tabular}{|c|c|c|c|}
 \hline
  Order & $\Lambda$/$\tilde{\Lambda}$  [MeV]& $\bar{d}_9$  [GeV$^{-2}$] & $L_2$  [GeV$^{-4}$]\\
  \hline
NLO & $400$/$500$ & $-0.010$ & $0.243$\\
NLO & $400$/$700$ & $-0.011$ & $0.249$\\
NLO & $550$/$500$ & $ 0.016$ & $0.605$\\
NLO & $550$/$600$ & $ 0.017$ & $0.731$\\
NLO & $550$/$700$ & $ 0.018$ & $0.892$\\
NNLO& $450$/$500$ & $-0.011$ & $0.188$\\
NNLO& $450$/$700$ & $-0.009$ & $0.173$\\
NNLO& $550$/$600$ & $ 0.005$ & $0.089$\\
NNLO& $600$/$500$ & $ 0.001$ & $0.113$\\
NNLO& $600$/$700$ & $-0.001$ & $0.028$\\
\hline
  \end{tabular}
 \caption{Values for $\bar d_9$ and $L_2$ found by fitting data up to $|{\bf q}|=400$ MeV, using different values of the cutoffs.}
\label{tab:LECs}
\end{table}

In Table~\ref{tab:LECs} we present the values of $\bar{d}_9$
and $L_2$ obtained in our fits. Small values of $\bar{d}_9$ are
preferred, which is consistent with the findings of
Ref.~\cite{Gasparyan:2010xz}. 
Reassuringly,  
 the inferred values of $\bar{d}_9$ show only a very mild dependence on
the cutoffs as compared to the expected natural size of this LEC, $|\bar{d_9}|
  \sim 1$ GeV$^{-2}$. In contrast, the values of $L_2$ do depend on
the choice of the regulator employed for the NN potential, as one
would expect. It is comforting to see that all obtained values of
$L_2$  are natural with respect to the cutoff scale $\Lambda$ employed
in these calculations. The values of $L_2$ reported in the table show
that  two-body effects in $G_M$
play a larger role in the calculation with NLO deuteron wave functions,
as seen in 
Fig.~\ref{fig:GMbreakdown}.

{\it Summary:} The first two-body effects in the deuteron magnetic form factor $G_M$, occur at $O(eP^4)$ in $\chi$EFT, i.e. three orders beyond leading. Inclusion of these
mechanisms in the computation of $G_M$ improves the description of
data, and allows exact reproduction of the deuteron magnetic moment,
which otherwise is underpredicted in $\chi$EFT. Experimental data is
then well described for $Q^2 < 0.35$ GeV$^2$, and the chiral expansion
for $G_M$ is found to converge well for $Q^2 < 0.25$ GeV$^2$, provided
that the NNLO wave functions of Ref.~\cite{Epelbaum:2004fk} are
employed. Finally, we note that the proposal of Ref.~\cite{NTvK05} to 
change the scaling of short-distance $\chi$EFT operators in order to ensure 
proper renormalization of the theory
does not significantly alter the relative importance of such operators in the ${}^3$S$_1$-${}^3$D$_1$ channel~\cite{Bi06}. 
Therefore we expect that the conclusions of this study
 will be quite robust with respect to developments on this front. 

\section*{Acknowledgements}
We thank Ulf-G. Mei\ss ner for useful comments on the ma\-nuscript. This work is supported by the EU HadronPhysics3 project
``Study of strongly interacting matter'',  
by the European Research Council (ERC-2010-StG 259218 NuclearEFT), 
by the DFG (TR 16, ``Subnuclear Structure of Matter''), and by the US
Department of Energy (contract no. DE-FG02-93ER40756).

\end{document}